\documentclass{article} 
\usepackage{iclr2021_conference,times}
\usepackage{authblk}

\usepackage{amsmath,amsfonts,bm}

\def\eqref#1{equation~\ref{#1}}

\def\1{\bm{1}}

\DeclareMathAlphabet{\mathsfit}{\encodingdefault}{\sfdefault}{m}{sl}
\SetMathAlphabet{\mathsfit}{bold}{\encodingdefault}{\sfdefault}{bx}{n}

\usepackage{hyperref}
\usepackage{url}
\usepackage{graphicx}
\usepackage{subcaption}
\usepackage{booktabs}

\title{Towards Interpreting Zoonotic Potential of Betacoronavirus Sequences With Attention}

\author[1,*]{\textbf{Kahini Wadhawan}}
\author[2,*]{\textbf{Payel Das}}
\author[3]{\textbf{Barbara A. Han}}
\author[3]{\textbf{Ilya R. Fischhoff}}
\author[3]{\textbf{Adrian C. Castellanos}}
\author[4]{\textbf{Arvind Varsani}}
\author[2]{\textbf{Kush R. Varshney}}
\affil[1]{IBM Research, New Delhi, India}
\affil[2]{IBM Research, Yorktown Heights, NY, USA}
\affil[3]{Cary Institute of Ecosystem Studies, NY, USA}
\affil[4]{The Biodesign Institute, Arizona State University, USA}

\affil[*]{kawadhaw@in.ibm.com and daspa@us.ibm.com}

\iclrfinalcopy 
\begin{document}

\maketitle

\begin{abstract}
Current methods for viral discovery target evolutionarily conserved proteins that accurately identify virus families but remain unable to distinguish the zoonotic potential of newly discovered viruses. Here, we apply an attention-enhanced long-short-term memory (LSTM) deep neural net classifier to a highly conserved viral protein target to predict zoonotic potential across betacoronaviruses. The classifier performs with a 94\% accuracy. Analysis and visualization of attention at the   sequence and structure-level features indicate possible association between important protein-protein interactions governing viral replication in zoonotic betacoronaviruses and zoonotic transmission.  
\end{abstract}

\section{Introduction}

The majority of viruses emerging in humans arise from animal hosts. For example, SARS-CoV-2, the virus causing the COVID-19 pandemic, belongs to the betacoronavirus($\beta$-CoVs) family from which numerous other viruses have been described from multiple mammal species around the world. The typical method for surveiling for such viruses is to identify a conserved region of the genome. For instance, the RdRp (RNA-dependent RNA polymerase) gene is a highly conserved sequence that is commonly used to ascertain $\beta$-CoVs presence from sampled wildlife species. While this conserved nature makes RdRp sequences a reliable target for accurately identifying novel $\beta$-CoVs from field surveillance, it does not enable predictions about whether the virus poses a zoonotic threat to humans. This is because the RdRp sequences encode the polymerase responsible for replicating the RNA genome once inside the cell, which is not involved in host cell entry or the infection process. The disconnect between broad viral surveillance and accurate predictions about zoonotic potential of newly discovered viruses remains a major research frontier, which has been highlighted by the emergence of SARS-CoV-2. Numerous closely-related $\beta$-CoVs have been identified from field surveillance over the past several years, but no clear and cost-effective way to triage the zoonotic risk posed by these newly discovered viruses. Beyond quantifying the zoonotic risk posed by a novel virus, preventing their emergence also relies on differentiating the key elements that enable some viruses to successfully infect humans. In other words, explaining what viral structures enable high zoonotic potential will be critical for informing what downstream actions will be effective for preventing spillover and viral emergence. 

Recent studies have shown stronger performance of machine learning based approaches for zoonotic transmission prediction task, when compared to homology based approaches, both at the protein sequence and gene sequence level \citep{eng2017predicting, xu2017predicting,li2018comparative, zhang2019rapid, mock2019viral,wang2013predicting, bartoszewicz2021interpretable}.  There is a consensus that deep models outperform shallow classifiers. 
Previous works have often used either feature vectors reflecting compositional and physicochemical properties of protein sequences or on word vectors.  In a few studies, input features were augmented with structural details, e.g. \citep{fischhoff2021predicting}. In this study, we develop a deep neural net classifier  for predicting zoonotic potential of RdRp sequences. For this purpose, we employ an attention mechanism enhanced long short-term memory (LSTM) model \citep{HochSchm97} trained on RdRp sequences, as attention \cite{bahdanau2014neural} is known to better cope with longer length sequences than a simple LSTM model, as is the case here. The final model yields a $94\%$ prediction accuracy. 

For wide-spread adoption of deep learning models for high-stake tasks, such as zoonotic transmission prediction, it is important that such models are interpretable and explainable.  
Towards this direction, we intend to interpret the attention signals of the intermediate layers of deep classifier by visualizing the  attention on the amino acid characters of the sample sequences. We also investigate average attention difference between two classes as a function of the amino acid type. Finally, we visualize the learned attentions on the 3D RdRp structure, as the 3D structure 
determines its function. Analysis of these internal workings of the classifier model provides important insights into likely mechanism of zoonotic transmission, as class-level attention differences map to the key protein-protein interaction sites. To our knowledge, this is the first machine learning work that provides sequence and structure-level understanding of zoonotic transmission potential of protein sequences. 

\vspace{-0.2cm}
\section{Data: RdRp Sequence Data and Label Annotation}
\label{gen_inst}
We assembled a dataset of published RdRp gene sequences for all $\beta$-CoVs, freely available in GenBank \citep{Benson2013-ye}.
For each published sequence, we added a binary label to designate zoonotic status based on the results of published research. The potential zoonotic sequences included ones similar or identical to sequences that caused MERS coronavirus (MERS-CoV) 
SARS. In the NCBI records, “SARS” was often used to describe these viruses; here we refer to SARS-CoV-1 to distinguish from SARS-CoV-2 causing the COVID-19 pandemic.  
MERS-CoV or SARS-CoV-1 sequences found in humans were categorized as zoonotic. If no host species was identified for a sequence, we did not categorize zoonotic status for that sequence. Prior to modeling, all sequences for which we were unable to categorize zoonotic status were removed from the dataset. For sequences passaged or adapted through cell culture of a different species than the original host,  we made no determination about zoonotic status, as passaging may have changed zoonotic capacity. These included viruses described as involving Vero cell culture, from African green monkey (\textit{Cercopithecus aethiops}) cells \citep{ammerman2008celllines, tsui2003coronavirusvariations}; mouse-adapted \citep{lu2009humoral}; recombinant; or LLC-MK2, cultured in rhesus monkey (\textit{Macaca mulatta}) cells (e.g., \citet{kaye2006cellreplication}). Information about these forms of cell culture was often found in the host field in NCBI; or as part of “passage details” in the notes field. If a sequence originated in one person and was cultured in cells from a different human cell line prior to sequencing, we considered it to remain zoonotic as we did not expect differences among people to change the zoonotic status.     

We also considered sequences found in animals to be zoonotic if the authors of the sequence considered it to be functionally identical to sequences found in people. For MERS-CoV, we emphasized virus taxonomic clade position and locality, which is related to clade position, as for MERS-CoV this information is more readily available than information about sequence similarity to human MERS-CoV. MERS-CoV in clades A and B include known human sequences and are known to cross from camels (\textit{Camelus dromedarius}) to humans \citep{lau2016mers, chan2014tropism}. For ambiguous MERS-CoV sequences, we categorized as zoonotic if there were human cases found in the same geographic area. MERS-CoV in Africa, in clade C, are similar but with no known human cases \cite{chu2018mers}. Clade C has lower titer and virulence in lab experiments, compared to clade A or B \cite{}. We assumed that clade C MERS-CoV to be non-zoonotic. 
For SARS-CoV-1 viruses found in animals, if insufficient information was available to confirm that the virus was zoonotic then it was labeled non-zoonotic. For example, SARS-CoV-1 viruses found in bats were categorized as non-zoonotic if no study was published associated with the sequence. SARS-CoV-1 reported in palm civet (\textit{Paguma larvata}) were considered zoonotic if the authors considered it to be identical to sequences found in people \citet{wang2005sars}, and otherwise categorized as non-zoonotic \citep{yi2005sars}. We considered as zoonotic a sequence isolated from an animal that also replicated in human cell lines, \textit{e.g.} a SARS-CoV-1 from \textit{Rhinolophus sinicus} bats that replicated in human cell culture \citep{ge2013isolation}. If the sequence was found in an animal species not known to be a bridge species to people, \textit{e.g.} a ferret, we categorized it as non-zoonotic \citep{town2012ferret}.

  

\section{Methods}
\label{headings}

The resulting dataset contained a total of 4259 sequences of length up to 910 characters, with vocabulary of 21 amino acids (A, C, D, E, F, G, H, I, K, L, M, N, P, Q, R, S, T, V, W, X, Y). We use 60/20/20 ratio to create train/val/test split, resulting in 2548, 876 and 835 sequences respectively, maintaining the positive to negative ratio in each split. A LSTM model enhanced with a single head attention mechanism was trained on this RdRp sequence dataset to predict their zoonotic potential (as a binary classification task). The model architecture was comprised of a single encoded layer with single head attention \cite{bahdanau2014neural}. The model accepts in the sequence of amino acids $x = (x_{1},x_{2},...,x_{L})$ and learns an embedding $z = (z_{1},z_{2},...,z_{L})$, which is passed to a dropout layer followed by a RELU and a softmax layer. Best test accuracy of $94\%$ was achieved after hyper-parameter search. More details are in SI section \ref{model_details_si}

\section{Results}
\begin{figure}[t]
\centering
  \begin{subfigure}[t]{0.26\textwidth}
    \centering
    \includegraphics[width=\textwidth]{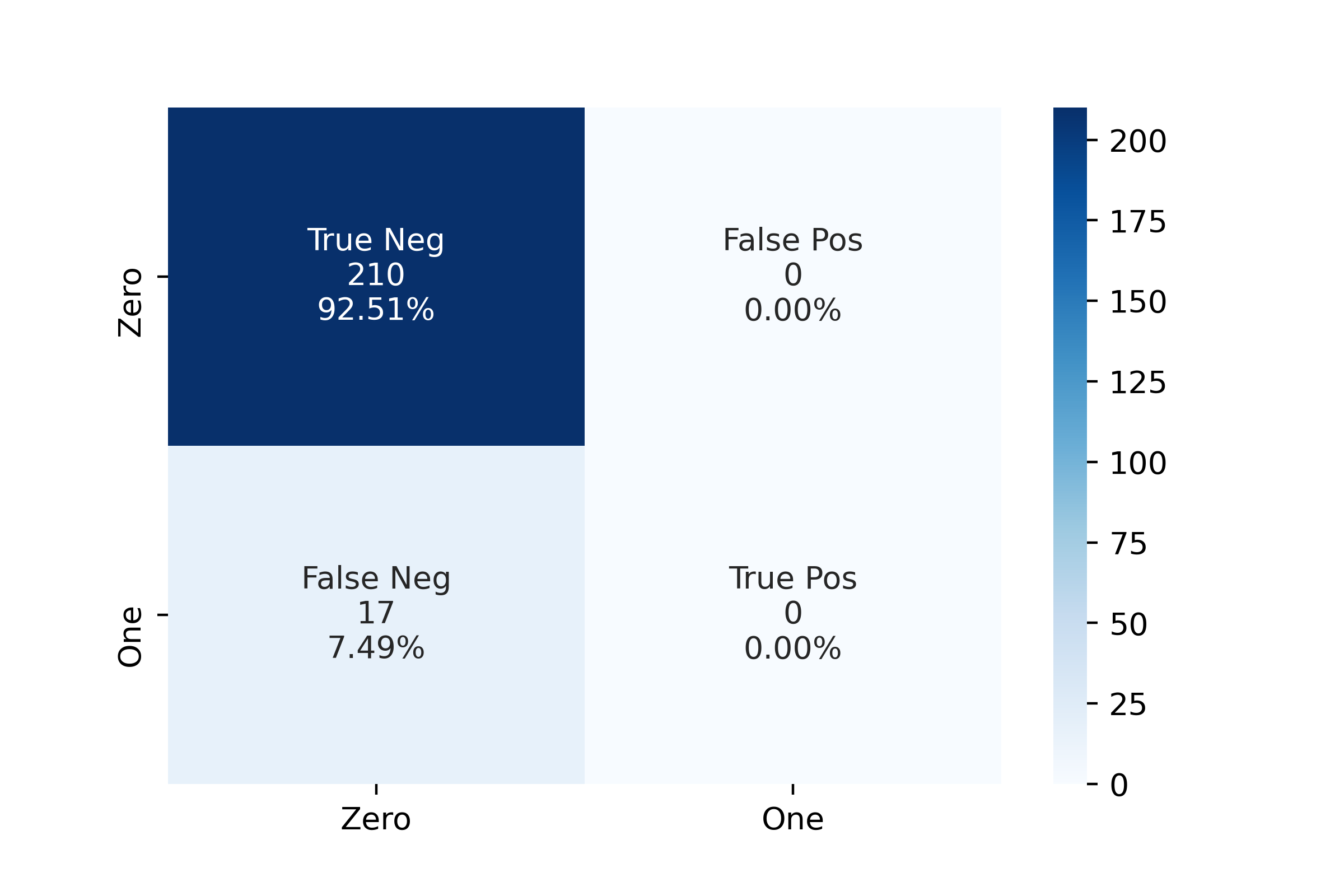}
    \caption{$len<150$}
    \label{fig:low_len_cfm}
  \end{subfigure}
      \hspace{-40cm}
  \hfill
  \begin{subfigure}[t]{0.26\textwidth}  
    \centering 
    \includegraphics[width=\textwidth]{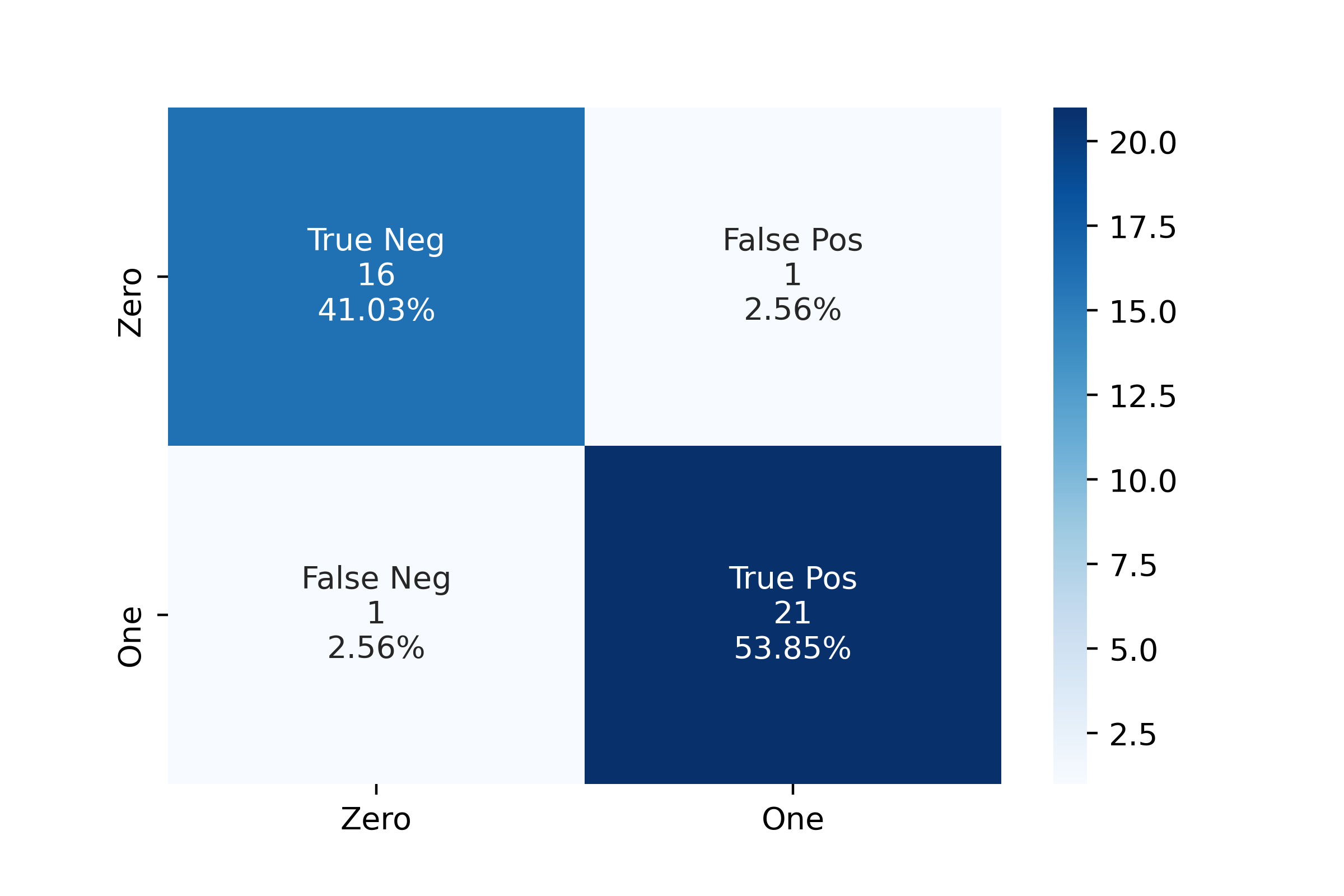}
    \caption{$150<len<500$}
    \label{fig:mid_len_cfm}
  \end{subfigure}
     \hspace{-40cm}
  \hfill
  \begin{subfigure}[t]{0.26\textwidth}  
    \centering 
    \includegraphics[width=\textwidth]{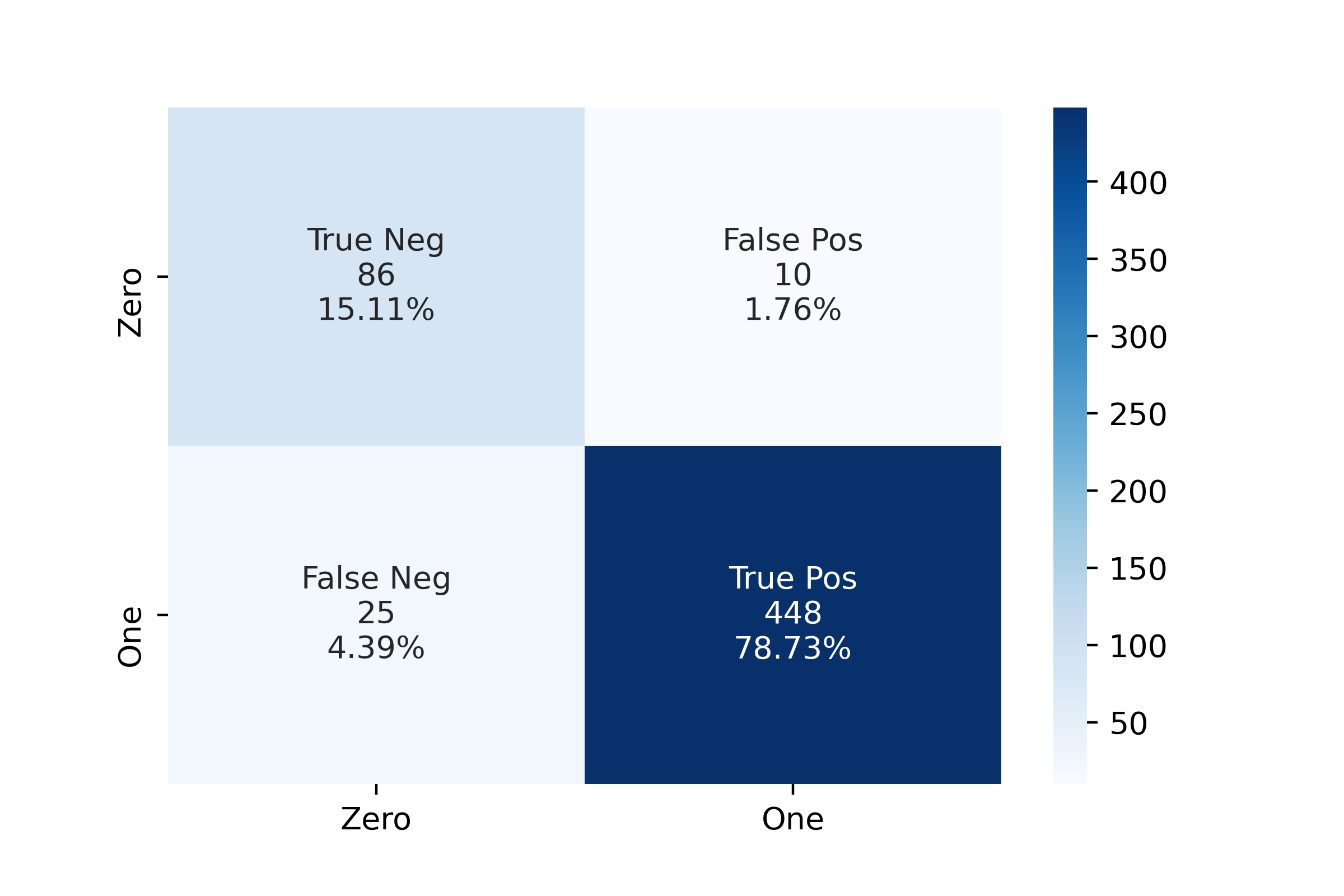}
    \caption{$len>500$}
    \label{fig:long_len_cfm}
  \end{subfigure}
     \hspace{-40cm}
  \hfill
  \begin{subfigure}[t]{0.26\textwidth}  
    \centering 
    \includegraphics[width=\textwidth]{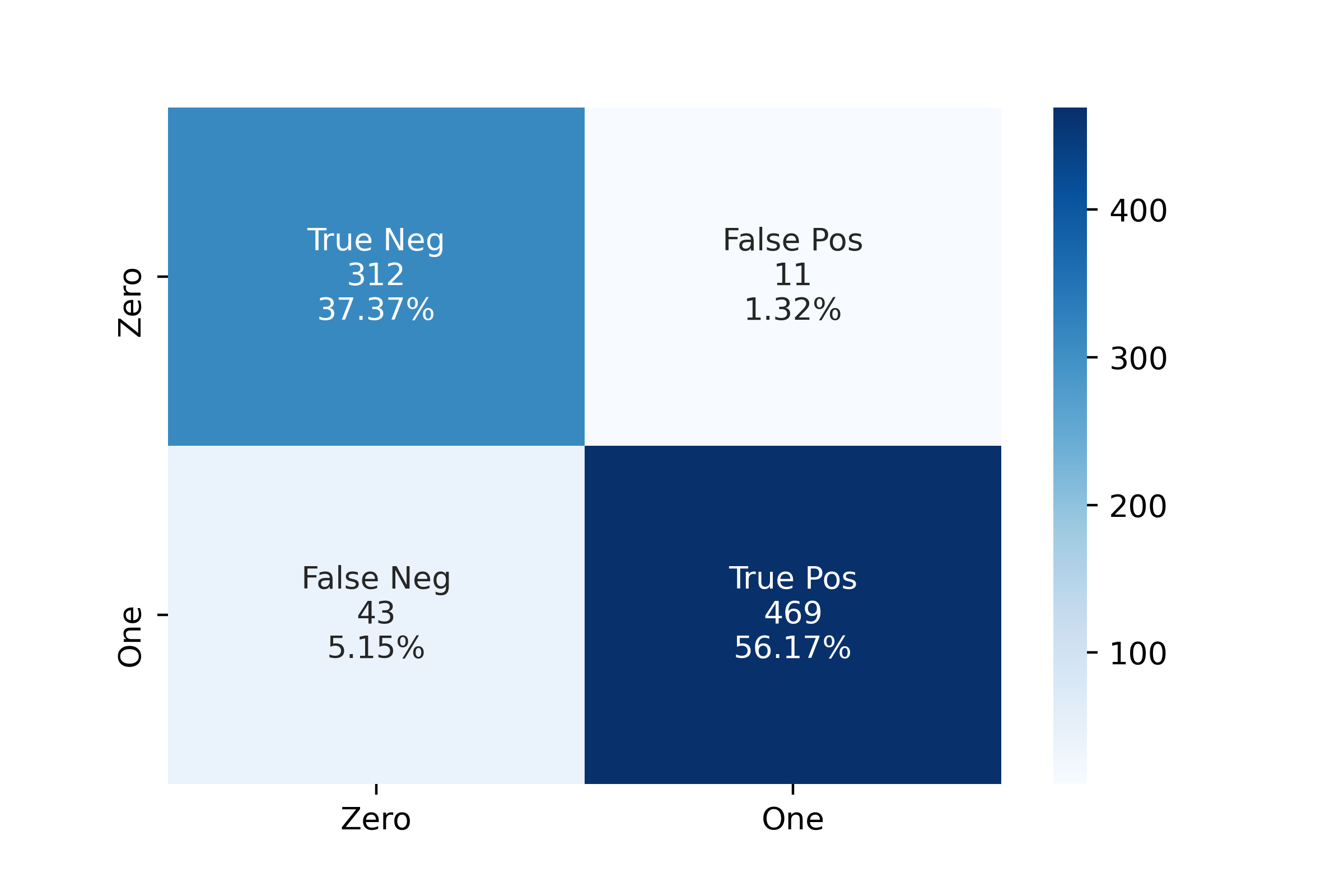}
    \caption{all length}
    \label{fig:all_len_cfm}
  \end{subfigure}
  \caption{Model performance confusion matrix computed over range of sequence lengths.} 
  \label{fig:len_cfm}
\end{figure}

\begin{figure}[t]
\centering
  \begin{subfigure}[t]{0.33\textwidth}
    \centering
    \includegraphics[width=\textwidth]{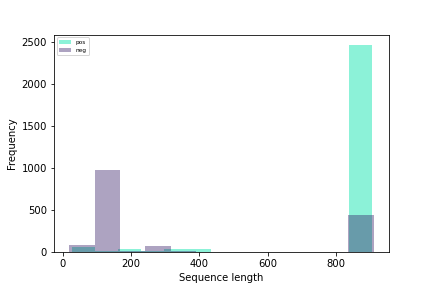}
    \caption{Length distribution}
    \label{fig:len_dist_data}
  \end{subfigure}
      \hspace{-40cm}
  \hfill
  \begin{subfigure}[t]{0.33\textwidth}  
    \centering 
    \includegraphics[width=\textwidth]{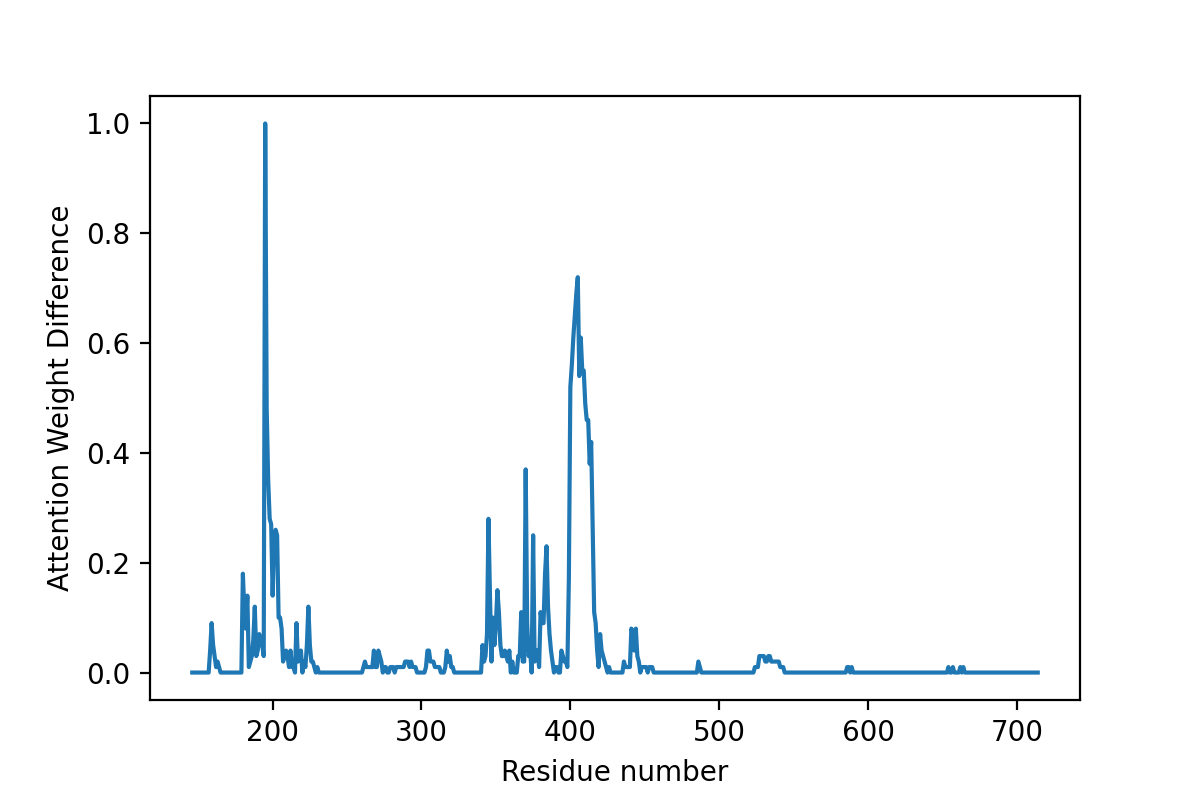}
    \caption{Avg. attention difference vs residue number}
    \label{fig:attn_diff_res}
  \end{subfigure}
     \hspace{-40cm}
  \hfill
  \begin{subfigure}[t]{0.33\textwidth}  
    \centering 
    \includegraphics[width=\textwidth]{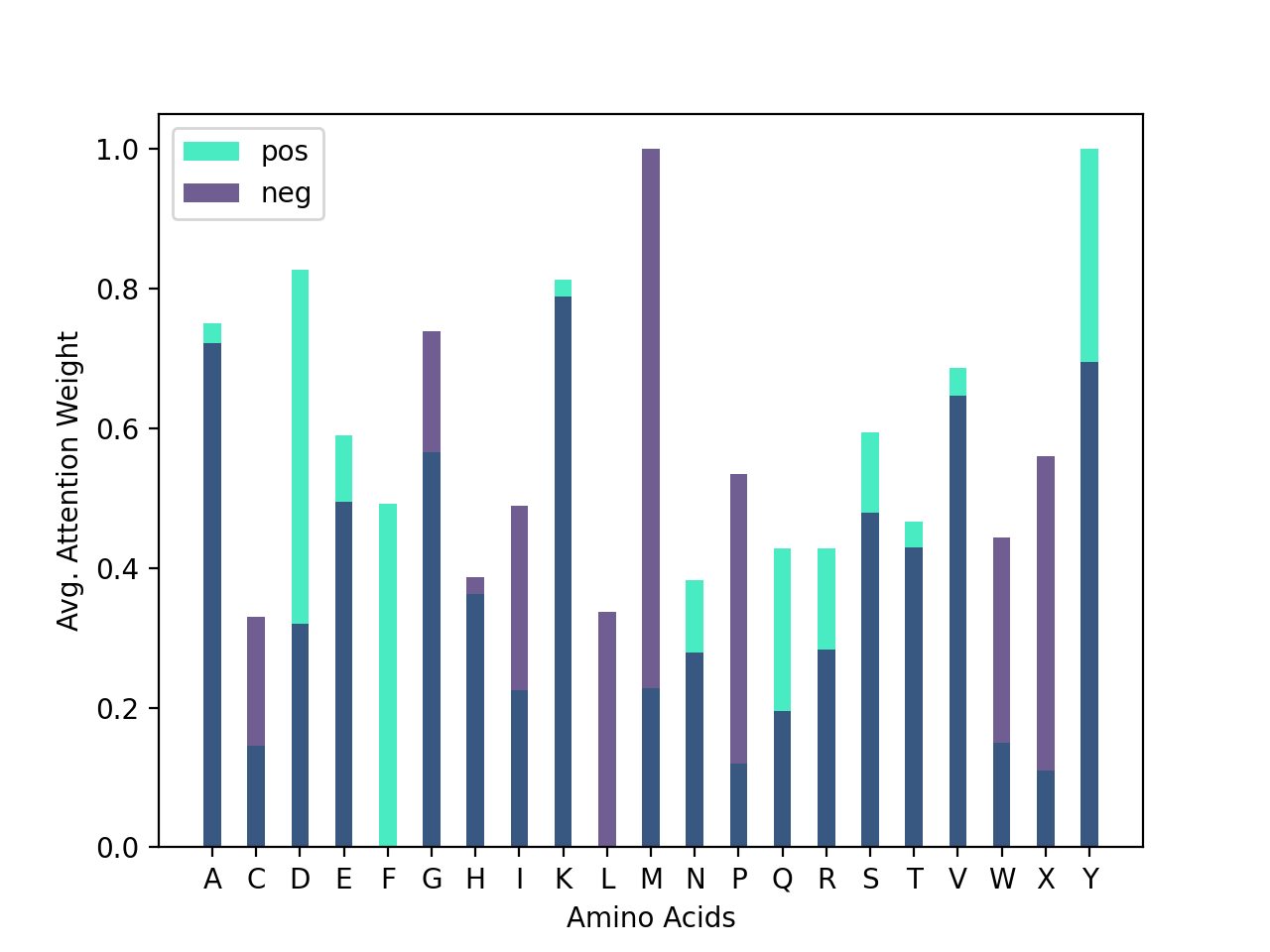}
    \caption{Feature importance (avg. attention over amino acids)}
    \label{fig:feat_imp}
  \end{subfigure}
   \caption{Data statistics and attention at 3D and 1D level plots.}
  \label{fig:data_and_1d_results}
\end{figure}

\textbf{Zoonotic classification accuracy.}
In this study, we use an attention-enhanced classifier, as attention has been proved to be helpful for dealing with longer sequences \citep{bahdanau2016neural}. In our dataset, the max sequence length is $910$. Figure \ref{fig:len_dist_data} shows the color coded length distribution of sequences for both positive and negative classes. The dataset is very length skewed. We see that the majority of zoonotic sequences are of length $>800$, while the shorter length $<200$ are mostly non-zoonotic. Despite the classification challenges presented by this skew, the best performing model achieved an overall  accuracy of $94\%$. Figure \ref{fig:len_cfm} shows confusion matrices for different length ranges. For shorter length sequences ($len < 150$), the classifier correctly predicted $> 92\%$ of the true negatives, but missed a small number of true positives. 
Improved predictions of true positives was achieved with increase in sequence length. F1-scores are reported in SI Table \ref{tab:f1-scores} and  section \ref{model_performance}.



\begin{figure}[htp]
    \centering
    \includegraphics[width=14cm]{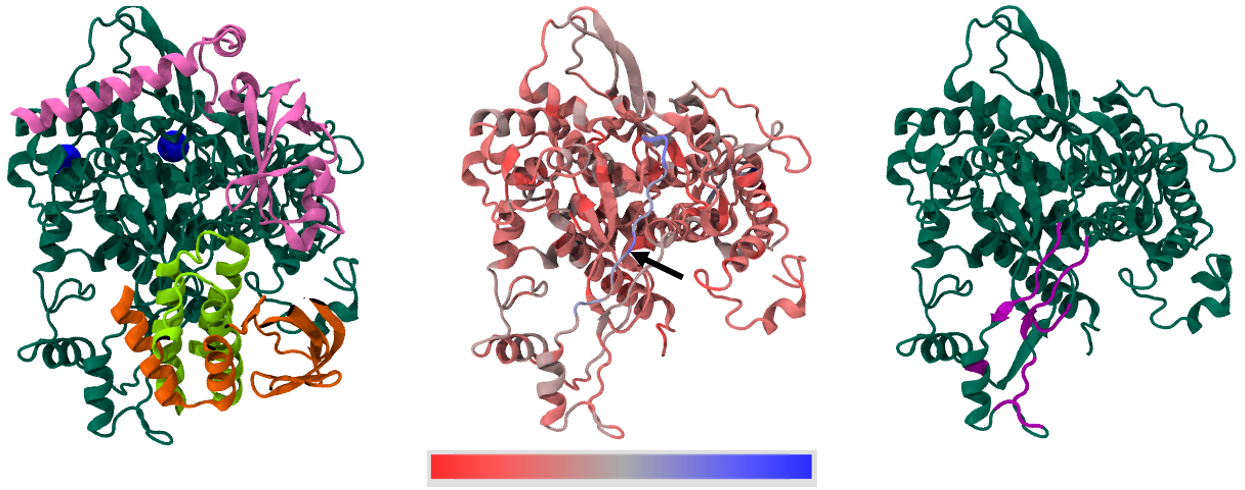}
    \caption{$Left:$ Structure of SARS-CoV nsp12 RdRp (dark green) bound to nsp7 (light green) and nsp8 (pink) co-factors (pdb id:6NUR). 
    Metal ions shown as blue spheres. $Middle:$ Average attention differences between two classes, mapped to the SARS-CoV nsp12 structure (6NUR.pdb chain A residue 146 to 714). Red-gray-blue (low to high) colorscale is used to visualize the attention heatmap. 
    $Right:$ Regions of RdRp interacting (cutoff: 6.5 \AA) with nsp7 highlighted in magenta.}
    \label{fig:3Dstructure}
\end{figure}

\textbf{Interpretation using Attention Visualization.}
Mapping predictions back to a target sequence has been used both as a means of  investigating a given model’s performance and as a method of sequence analysis. For example, convolutional filter visualization has been used for genome sequence classification \citep{bartoszewicz2021interpretable}. Protein sequence position importance was investigated by an attention function in subcellular localization classification task \citep{jurtz2017introduction}. Recently, 3D visualization of attention of a Transformer protein sequence model revealed accurate capture of structural and binding properties \citep{vig2020bertology}. Similarly, in this study, we visualize and identify regions of 1D protein sequence and 3D protein structure that are important for the prediction of zoonotic potential.
Figure \ref{fig:feat_imp} shows averaged normalized attentions for each amino acid character, estimated by averaging over all sequences within the same class. At the amino acid level, significant differences in attention are evident between two classes. For example, relatively higher attention in positive sequences is found in the amino acids D, Y, R, F, and Q. 

Since the 3D structure of the protein determines its function, we project the average attention heatmap differences between the two classes on the publicly available SARS-CoV nsp12 RdRp structure  bound to the partnering nsp7 and nsp12 cofactor proteins \citep{kirchdoerfer2019structure}(Figure  \ref{fig:3Dstructure}). Figure \ref{fig:attn_diff_res} show these attention differences as a function of residue number of nsp12 template. For selection criteria refer SI section \ref{criteria_si}. Figure \ref{fig:feat_imp_long_3d_single_long} further highlights these differences mapped on to the homology model (using SARS-CoV nsp12 as template) of individual sequences from the positive and negative class. These visualizations clearly show higher attentions within the positive sequences map to the N-terminal structural regions that correspond to the nsp12-nsp7 interaction surface. This result is interesting in light of prior work showing that the assembly of RdRp with cofactor proteins is crucial for polymerase activity that governs viral replication \citep{yin2020structural,subissi2014one}. In fact, protein-protein interaction design has been shown as a proven  approach for RdRp inhibition in other related viruses
\citep{subissi2014one}. 



\section{Discussion and Future Work}
We present an attention-enhanced LSTM deep neural net classifier that predicts zoonotic transmission in $\beta$-CoVs. Crucially, this classification was achieved using widely available RdRp sequences. 
Accurate prediction of zoonotic potential from RdRp sequences in $\beta$-CoVs (or similarly used reliable primer sequences for other viral families) has the potential to alleviate a critical bottleneck in viral surveillance, which has so far been unable to infer zoonotic potential of novel viruses discovered as part of routine surveillance. 
Further investigation of attention maps at both the sequence and the structure level provided important interpretability of the ``black-box'' model. Our results indicate mapping of class-level attention differences at protein-protein interaction sites of the RdRp structure. Those sites are of significant functional importance for binding with cofactors crucial for polymerase activity involved in viral replication. Thus, while RdRp itself is not directly involved in the infection process, physiochemical interactions with neighbouring cofactors appear to influence the function of zoonotic $\beta$-CoVs compared to non-zoonotic counterparts. Future work will explore more sophisticated attention mechanism; multi-head (e.g. Transformers \citep{devlin-etal-2019-bert} \citep{vaswani2017attention}) for better interpretability and prediction. We also plan extend the current study beyond $\beta$-CoVs to investigate the robustness in  capturing of biological context via  model features.  

\bibliography{iclr2021_conference}
\bibliographystyle{iclr2021_conference}

\appendix
\section{Appendix}
\subsection{Model Details}
\label{model_details_si}
\subsubsection{The Model}
For sequential data, Recurrent Neural Networks were introduced but as the sequence grows, the RNNs, weights could grow beyond control or vanish. Long Short-Term Memory \citep{HochSchm97} was proposed to deal with vanishing gradient problem \citep{Hochreiter:01book} and learn long-term dependency among longer time period. Attention further learns these dependencies in a smarter and relevant way \citep{luong2015effective, Galassi_2020, bahdanau2014neural}. 

LSTM estimates conditional probability $p(y_{
1},...,y_{T}|x_{1},...,x_{L})$. Our task was binary classification, so the input was sequence of characters ($x_{1},...,x_{L}$) and output was binary probability vector among two classes, positive and negative such that $\sum_{t=1}^{2} p(y_{t})=1$. The LSTM computes this conditional probability by first obtaining the fixed-dimensional representation $h$ of the input sequence and given by the last hidden state of the LSTM. We enhanced our model with attention so, a final hidden attentional state $h'_{t}$ was computed which was passed to a dropout layer, RELU and followed by softmax to obtain predictive distribution over classes. 


For the attention part, we used multiplicative attention \citep{luong2015effective}. All the hidden states of the encoder LSTM $h_{t}$ were used to compute context vector $c_{t}$. A variable length alignment vector was computed using current target hidden state $h_{t}$ with each source hidden state $h_{s}$

\begin{equation}
\label{attn_1}
     \alpha_{t}(s) = align(h_{t}, h_{s}) = \frac{\exp(score(h_{t}, h_{s}))}{\sum_{s'} \exp(score(h_{t}, h_{s'})}
\end{equation}

Here, $score$ is content-based function for which we used dot product, refer  \citep{luong2015effective} for other alternatives. 

\begin{equation}
\label{attn_2}
     score(h_{t}, h_{s}) = h_{t}^T h_{s}
\end{equation}

Given the alignment vector as weights, the context vector $c_{t}$ is computed as the weighted average over all the source hidden states.\citep{luong2015effective}. 

We took the dot product of the last hidden state as the target $h_{t}$ and context vector $c_{t}$ produce attentional hidden state. This attentional vector $h'_{t}$ was then passed to softmax layer get predictive distribution over the positive and negative class: 

\begin{equation}
\label{attn_3}
     p(y_{t}|y<t,x) = softmax(W_{s}, h'_{t})
\end{equation}







\subsubsection{Training Details}
\label{model_training}
We used a unidirectional LSTM with single layer and single head attention with 1000 cells (max sequence length) and 200 dimensional word embeddings (amino acid characters in our case) with a vocabulary of 21 amino acids. Hyper-parameter search was conducted and model giving best accuracy was found with the configuration of input embedding layer of size 200, lstm hidden layer of size 200, dropout 0.2, trained for 20 epochs. The baseline LSTM model with no attention gave $85\%$ accuracy.

\subsubsection{Model Performance Improvements on balanced Dataset}
\label{model_performance}
\begin{figure}[t]
\centering
  \begin{subfigure}[t]{0.26\textwidth}
    \centering
    \includegraphics[width=\textwidth]{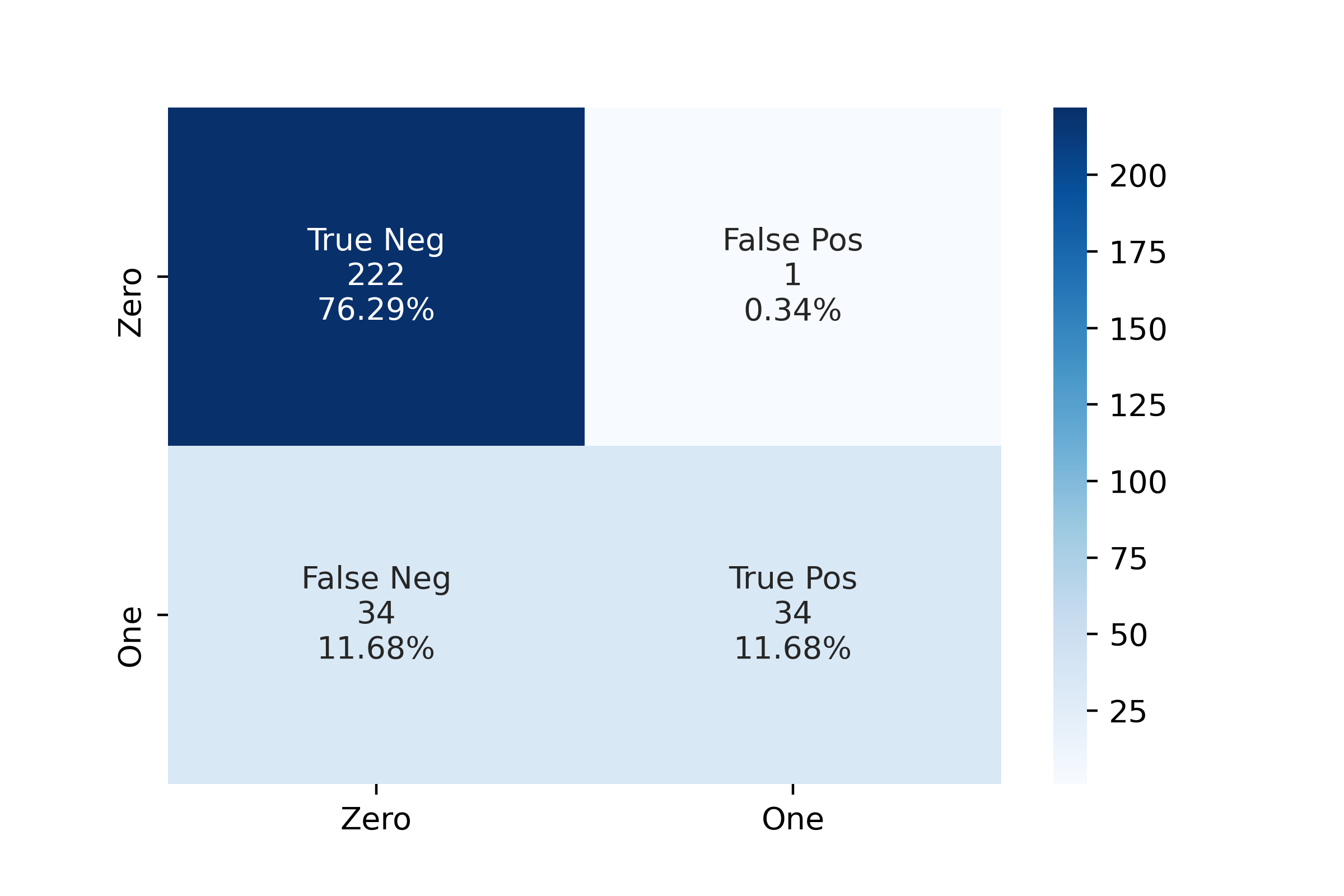}
    \caption{$len<150$}
    \label{fig:low_len_cfm_bal}
  \end{subfigure}
      \hspace{-40cm}
  \hfill
  \begin{subfigure}[t]{0.26\textwidth}  
    \centering 
    \includegraphics[width=\textwidth]{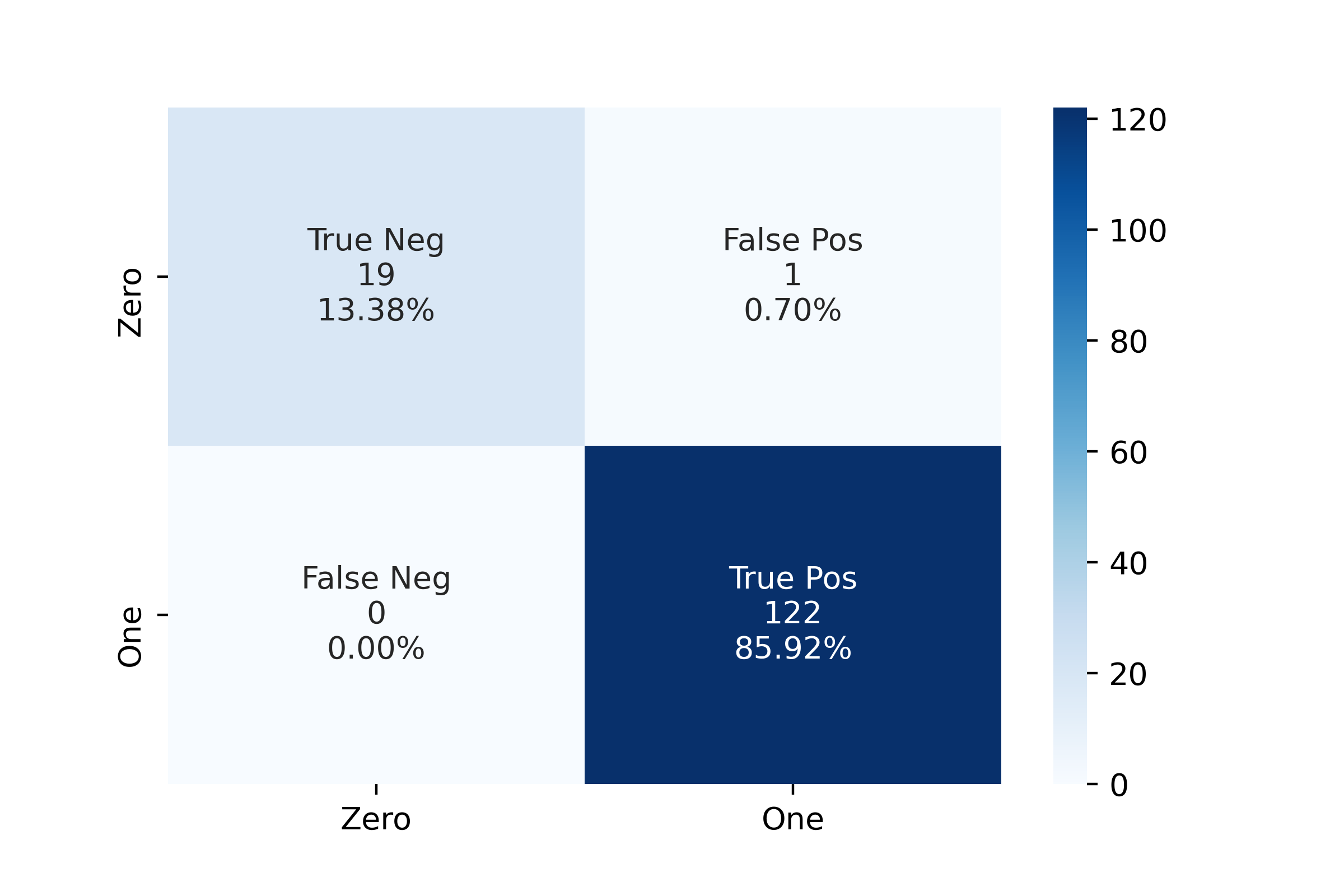}
    \caption{$150<len<500$}
    \label{fig:mid_len_cfm_bal}
  \end{subfigure}
     \hspace{-40cm}
  \hfill
  \begin{subfigure}[t]{0.26\textwidth}  
    \centering 
    \includegraphics[width=\textwidth]{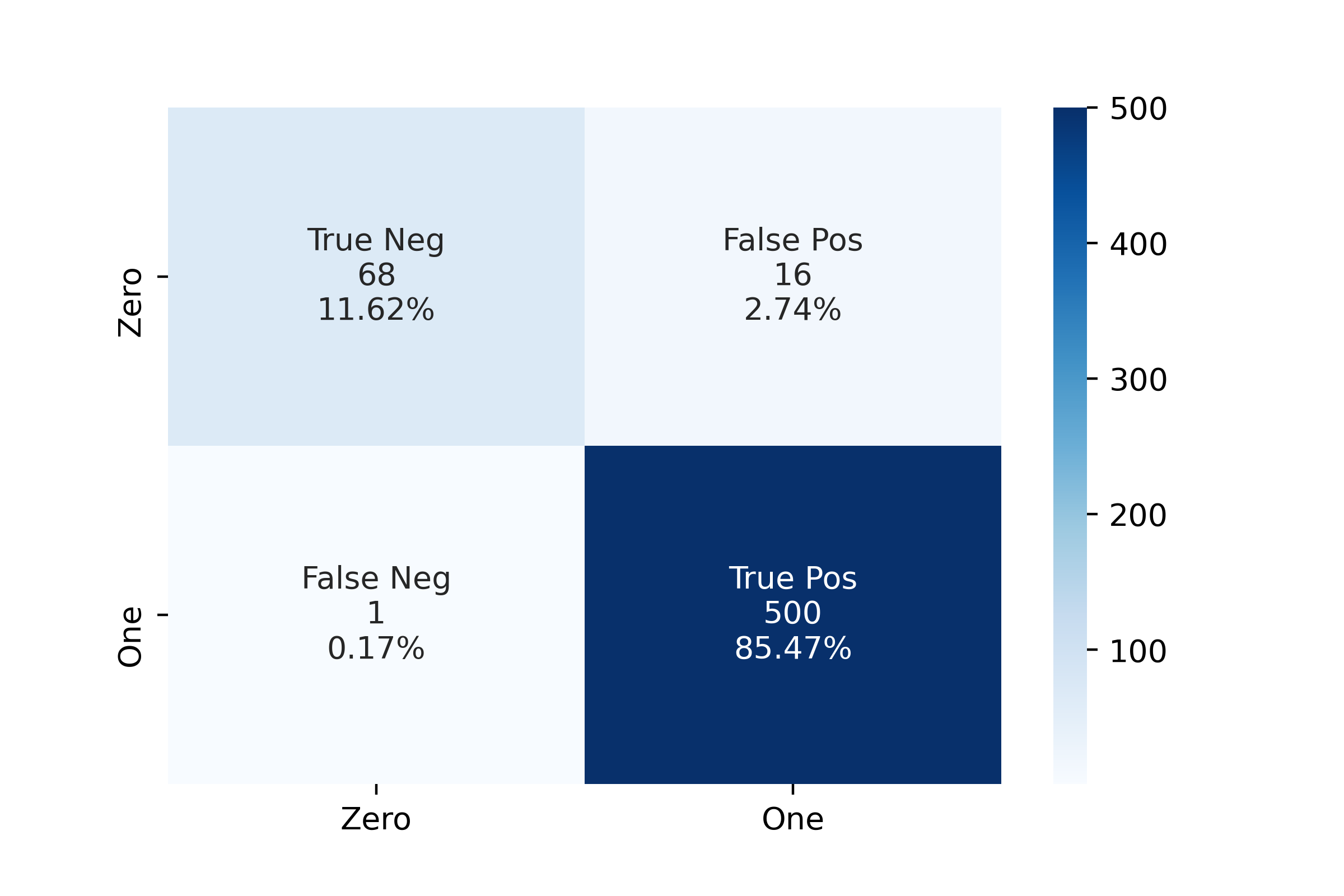}
    \caption{$len>500$}
    \label{fig:long_len_cfm_bal}
  \end{subfigure}
     \hspace{-40cm}
  \hfill
  \begin{subfigure}[t]{0.26\textwidth}  
    \centering 
    \includegraphics[width=\textwidth]{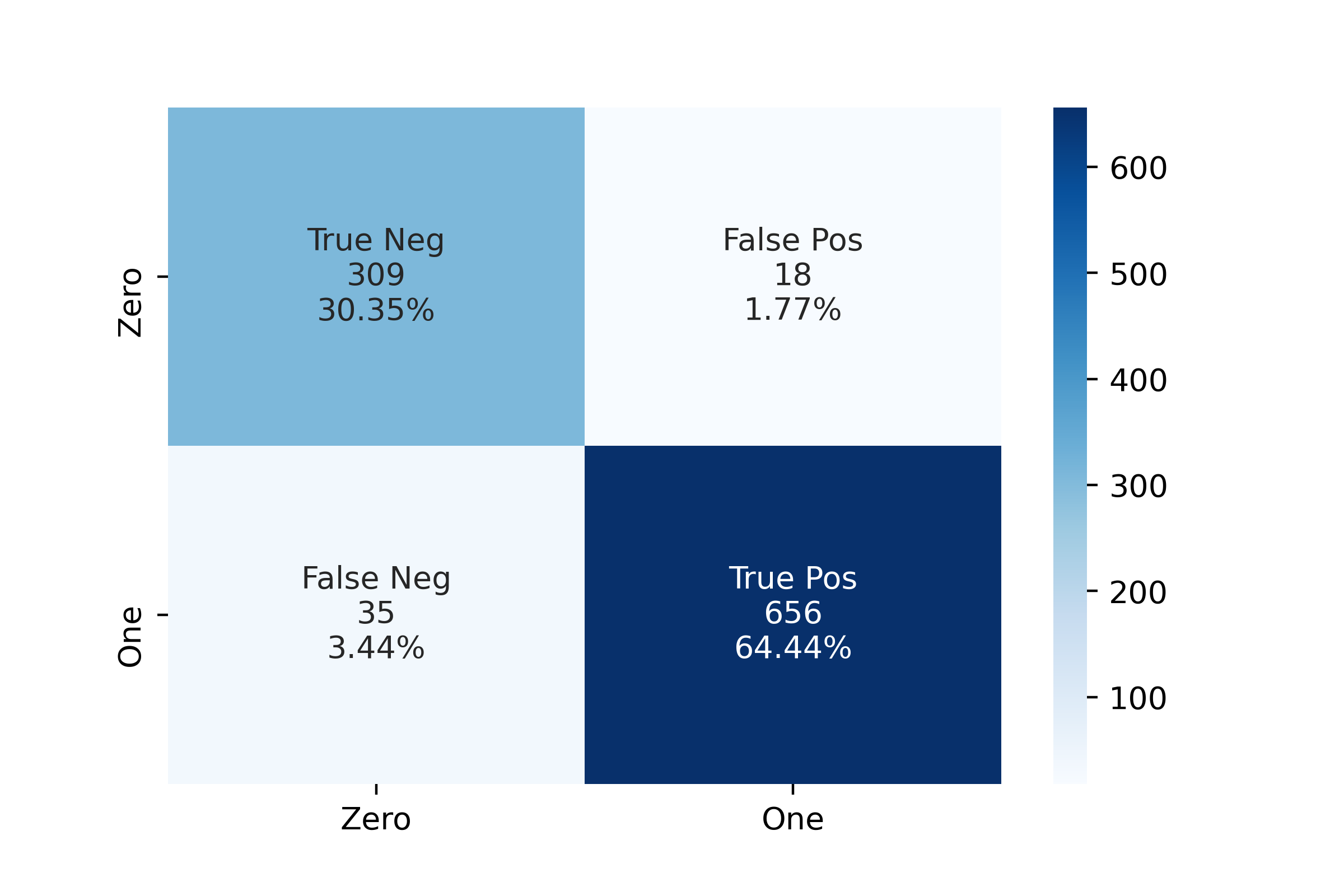}
    \caption{all length}
    \label{fig:all_len_cfm_bal}
  \end{subfigure}
  \caption{Model performance confusion matrix computed over range of sequence lengths for balanced dataset refer SI section \ref{model_performance}. 
  } 
  \label{fig:len_cfm_balanced}
\end{figure}

Due to the imbalanced class distribution as a function of sequence length, model performs poorly in lower sequence length regime len $<150$. It seems to overfit to negative class and hence we saw $0$ True positives in len $len<150$ fig \ref{fig:len_cfm}. Also, relatively poor performance with mid-length range. To overcome this problem, we adopted techniques to balance our dataset especially in lower length regime by oversampling the positive samples. In the imbalanced dataset we had around $195$ positive sequences with len $<400$ compared to $~1100$ negative class sequences (refer figure \ref{fig:len_dist_data}). So, we made 5 copies of each positive sample with length $<400$ which balances the dataset in lower length regime. Model was retrained on balanced data using configuration reported in SI section \ref{model_training} 

This improved the F1 score significantly (from 0 to 0.660) in the lower length regime. Improvements in recall and F1-scores of lower and mid-length ranges can be seen in table \ref{tab:f1-scores}. The overall F1-score of the model for all length also gets improved from $0.946$ to $0.961$. Figure \ref{fig:len_cfm_balanced} shows confusion matrices of model performance trained on balanced dataset. In the future work, we will explore more techniques to balance the dataset. 


\begin{table}[ht]
    \centering
    \begin{tabular}{p{4.0cm}|p{1.4cm}|p{1.4cm}|p{1.4cm}|p{1.4cm}}
        \toprule
        Sequence length ranges & Accuracy &  Precision & Recall
        & F1-score\\ 
        \hline   
        len$<150$ (\textbf{\textit{imbalanced data}})  & 0.925 & \textbf{NaN(0/0)} & \textbf{0.0} & \textbf{0.0} \\
        $150<$len$<500$ & 0.949 & 0.955 & 0.955 & 0.955 \\
       len$>500$ & 0.938 & 0.978 & 0.947 & 0.962 \\
        all length & 0.935 & 0.977 & 0.916 & \textbf{0.946}\\ 
        \hline
        len$<150$ (\textbf{\textit{balanced data}})  & 0.880 & \textbf{0.971} & \textbf{0.500} & \textbf{0.660}\\
        $150<$len$<500$  & 0.993 & 0.992 & 1.00 & 0.996 \\ 
        len$>500$  & 0.971  & 0.969 & 0.998 & 0.983 \\ 
        all length  & 0.948 & 0.973 & 0.949 & \textbf{0.961} \\ 
         \bottomrule
    \end{tabular}
    \caption{Model performance report using accuracy, precision, recall and F1-score for both imbalanced skewed dataset and balanced (oversampled positive datapoints) dataset over all sequence length ranges shows improvement in model performance by improvement in F1-scores in balanced dataset. Lower length regime ($len<150$) shows significant increase in F1-score. Interesting significant improvements are highlighted by bold text. Precision is not defined(\textbf{NaN}) in len$<150$ for imbalanced data and False negatives $> 0$ in this case so we take F1 to be 0 as model failed to identify any True positives when it could have done. }
    \label{tab:f1-scores}
\end{table}

\subsection{Attention Analysis Methods}
For interpreting the model, we carried different analysis on the attention vectors. Visualization of average attention vectors looked into the overall trends distributed into positive and negative classes at the structure level. In addition, we also investigated attention at each sample/individual level. 

\subsubsection{Sequence Selection Criteria for Average Attention Analysis}
\label{criteria_si}

We used attention vectors of the trained LSTM model and carried out detailed analysis on them. We computed average attention (shown in figure \ref{fig:3Dstructure}) of filtered positive (zoonotic) and negative (non-zoonotic) classes, to have an insight into what model learns about structural differences between the two classes. For mapping to a template structure, positive and negative sequences that are highly similar to SARS-CoV nsp12 RdRp structure bound to nsp7 and nsp8 co-factors (pdbid:6NUR) were selected. 

Sequence similarity between RdRP sequences and fixed target 6NUR was computed using NCBI BLAST similarity search \citep{blast}. Command used is 'blastp' using an evalue of 10, word size of 3, the BLOSUM62 subsitution matrix, a gap open penalty of 9, a gap extension penalty of 1, threshold of 16, comp based stats set to 0 and window size of 15.
Sequences with $alignment$ $len > 900$, $identities > 90\%$, $gaps=0$ and predicted correctly (true positives and true negatives) by our classifier model are selected. 277 positive and 26 negative sequences are chosen for this analysis shown in figures \ref{fig:3Dstructure} and \ref{fig:attn_diff_res}.

\subsubsection{Homology Modeling for individual level attention analysis}
\label{analysis_methods_si}
\begin{figure}[t]
\centering
  \begin{subfigure}[t]{0.48\textwidth}
    \centering
    \includegraphics[width=\textwidth]{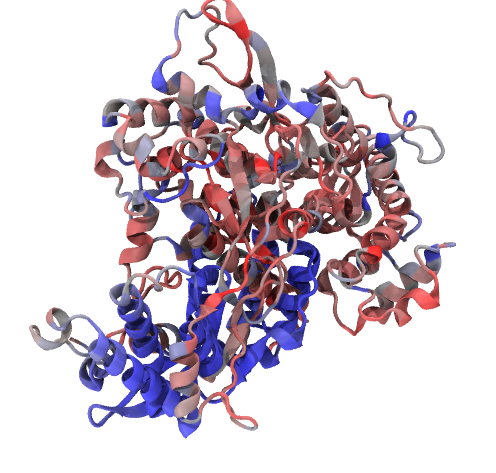}
    \caption{Positive (zoonotic) sequence}
    \label{fig:pos_feat_imp}
  \end{subfigure}
      \hspace{-40cm}
  \hfill
  \begin{subfigure}[t]{0.48\textwidth}  
    \centering 
    \includegraphics[width=\textwidth]{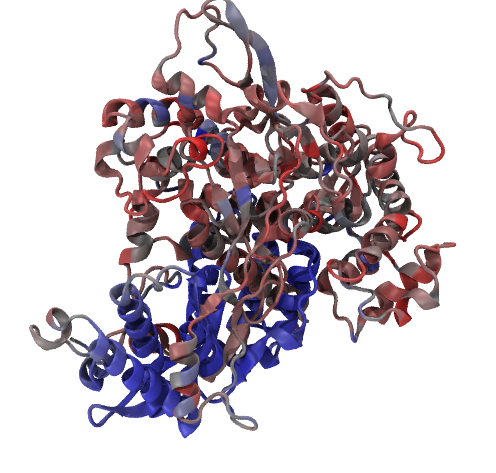}
    \caption{Negative (non-zoonotic) sequence}
    \label{fig:neg_feat_imp}
  \end{subfigure}
  \caption{Attention heatmap of one positive and one negative sequences with length $> 900$ mapped on to their modeled structure obtained using homology modeling (6NUR.pdb chain A used as template). Colorscale used: red-gray-blue (low to high)} 
  \label{fig:feat_imp_long_3d_single_long}
\end{figure}

\begin{figure}[t]
\centering
  \begin{subfigure}[t]{0.48\textwidth}
    \centering
    \includegraphics[width=\textwidth]{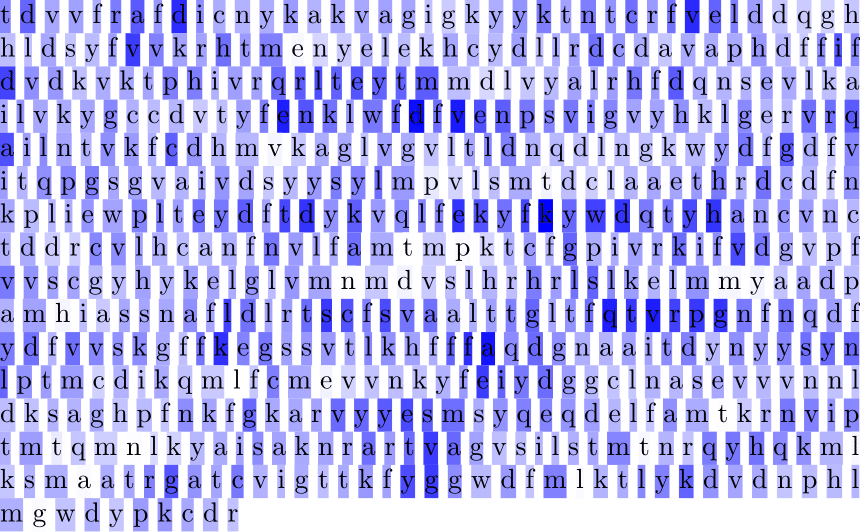}
    \caption{Positive(zoonotic) sequence, blast identities$>70\%$, classifier confidence score$>90\%$}
    \label{fig:pos_long_len_1d}
  \end{subfigure}
      \hspace{-40cm}
  \hfill
  \begin{subfigure}[t]{0.48\textwidth}  
    \centering 
    \includegraphics[width=\textwidth]{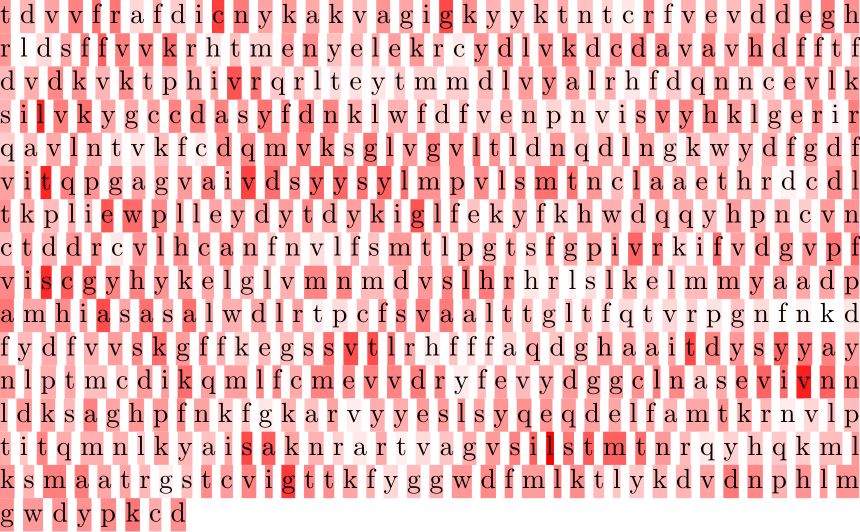}
    \caption{Negative(non-zoonotic) sequence, blast identities$>70\%$, classifier confidence score$>90\%$}
    \label{fig:neg_long_len_1d}
  \end{subfigure}
  \caption{Attention heatmap of one positive and one negative sequences with length $> 900$ at sequence 1D level. Positive with blue and negative with red. } 
  \label{fig:pair_long_len_seq_attn_1d}
\end{figure}

\begin{figure}[t]
\centering
  \begin{subfigure}[t]{0.48\textwidth}
    \centering
    \includegraphics[width=\textwidth]{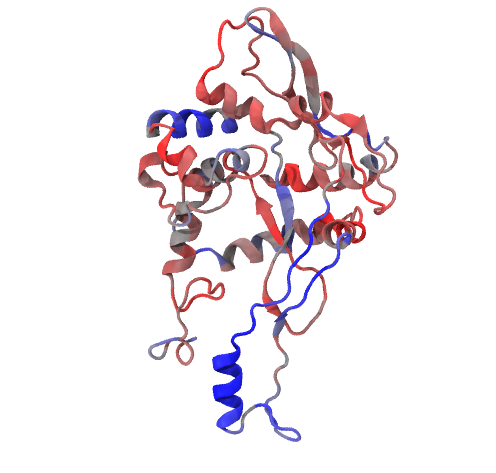}
    \caption{Positive(zoonotic) sequence len$~300$, blast identities$>70\%$, classifier confidence score$>90\%$}
    \label{fig:pos_short_len_seq_attn}
  \end{subfigure}
      \hspace{-40cm}
  \hfill
  \begin{subfigure}[t]{0.48\textwidth}  
    \centering 
    \includegraphics[width=\textwidth]{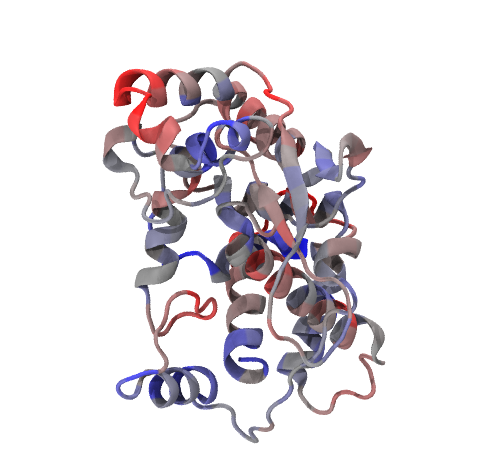}
    \caption{Negative(non-zoonotic) sequence len$~300$, blast identities$>70\%$, classifier confidence score$>90\%$}
    \label{fig:neg_short_len_seq_attn}
  \end{subfigure}
  \caption{3D attention heatmap, a pair of shorter length sequences that have high similarity (identities $>70\%$) with NSP12 chain A. For shorter length, used relaxed identities criteria as the overall length reduced so no sequences available with higher $90\%$ identities. } 
  \label{fig:pair_short_len_seq_attn}
\end{figure}

\begin{figure}[t]
\centering
  \begin{subfigure}[t]{0.48\textwidth}
    \centering
    \includegraphics[width=\textwidth]{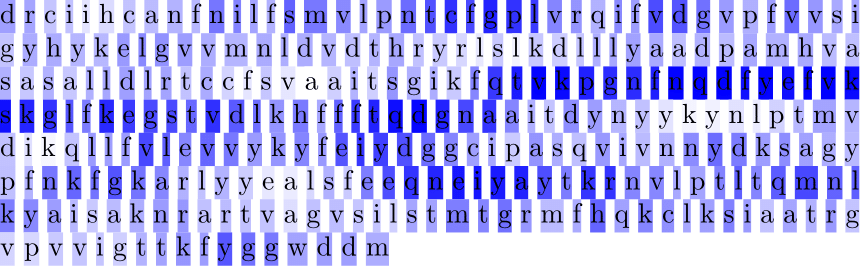}
    \caption{Positive(zoonotic) sequence len$~300$, blast identities$>70\%$, classifier confidence score$>97\%$}
    \label{fig:pos_short_len_seq_attn_1d}
  \end{subfigure}
      \hspace{-40cm}
  \hfill
  \begin{subfigure}[t]{0.48\textwidth}  
    \centering 
    \includegraphics[width=\textwidth]{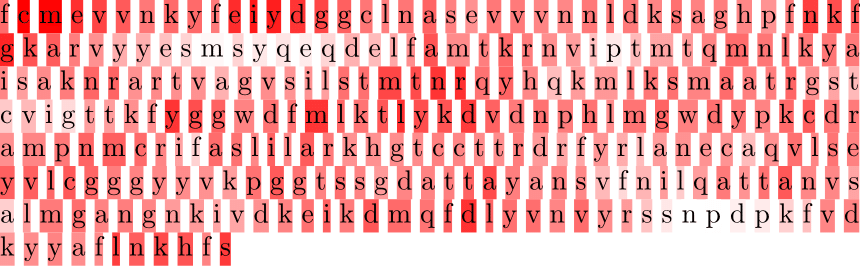}
    \caption{Negative(non-zoonotic) sequence len$~300$, blast identities$>70\%$, classifier confidence score$>97\%$}
    \label{fig:neg_short_len_seq_attn_1d}
  \end{subfigure}
  \caption{Attention heatmap of one positive and one negative sequences with length $~300$ at sequence 1D level. Positive with blue and negative with red.} 
  \label{fig:pair_short_len_seq_attn_1d}
\end{figure}

For visualizing attention at the structural (3D) level of individual positive and negative sequences (shown in figure \ref{fig:feat_imp_long_3d_single_long}), their structure prediction was required as the structures for these surveilled RdRp sequences were not available in the Protein Data Bank(PDB \citep{10.1093/nar/28.1.235}). For the same, homology modeling was conducted using SWISS-MODEL server \citep{10.1093/nar/gky427} using NSP12 Chain A as the template. Figure \ref{fig:3Dstructure},  \ref{fig:pair_long_len_seq_attn_1d},  \ref{fig:pair_short_len_seq_attn} and \ref{fig:pair_short_len_seq_attn_1d}  display attention visualization on structure and sequence level for pair of true positive and true negative long length and short length sequences.

Highly homologous sequences were filtered using $identities > 70\%$ and $gaps=0$. 
Alignment length $> 900$ gave longer length selections and for shorter length we set alignment length between $200$ to $300$. We employed homology modeling of these highly similar sequences using Swiss-model server.


\end{document}